# Brownian soliton motion

Yaroslav V. Kartashov,[1] Victor A. Vysloukh,[2] and Lluis Torner[1]

[1]*ICFO-Institut de Ciencies Fotoniques, Mediterranean Technology Park, and Universitat Politecnica de Catalunya, 08860 Castelldefels (Barcelona), Spain*

[2]*Departamento de Actuaria, Fisica y Matematicas, Universidad de las Americas – Puebla, Santa Catarina Martir, 72820 Puebla, Cholula, Mexico*

We reveal fundamental analogies between soliton dynamics in light-induced random photonic lattices and Brownian motion of particles. In particular, we discover that the average squared soliton displacement increases linearly with distance after an initial ballistic regime of propagation. We also find that in shallow lattices the average soliton displacement grows linearly with increasing lattice depth.

  Almost two centuries ago the botanist Robert Brown discovered that small particles suspended in liquids move chaotically. Later Albert Einstein predicted that the motion is caused by particle collisions with surrounding molecules. The intensity of this Brownian motion was found to increase with temperature, since the mean kinetic energy $3kT/2$ ($T$ is the temperature) of the liquid molecules equals the mean kinetic energy $m\langle v^2 \rangle/2$ of the suspended particles [1,2]. By changing randomly the direction of motion and its velocity upon elastic collisions with liquid molecules, a Brownian particle moves away along complex paths with time $t$, with a mean-square displacement given by $\langle r^2 \rangle \sim kTBt$, where the coefficient $B$ stands for the particle mobility.

  Here we are concerned with a similar physical setting, defined by nonlinear localized wave excitations evolving in a random potential made by an optical lattice. Analogies between nonlinear excitations and particles are known to occur in materials that support soliton propagation (see Ref. [3]). Soliton motion in materials with inhomogeneous transverse properties mimics the random motion of mechanical particles. In this context, optical induction techniques [4-7] opened up a wealth of opportunities to the generation of reconfigurable refractive index landscapes. Regular, perfectly periodic or modulated lattices can be used for the controllable steering and dragging of solitons. Optical induction allowed the recent observation of Anderson localization with linear light beams propagating in disordered optical lattices [8]. The interplay between disorder and nonlinearity has been extensively studied in different systems, including systems with random-point impurities [9,10], and one- [11], and two-dimensional [12] discrete waveguide arrays (see also reviews [13]). Disorder may result in localization of walking solitons [14] or, vice versa, it may facilitate the soliton transport [15]. In this Communication we reveal new analogies between the dynamics of strongly nonlinear excitations in optically induced random lattices and Brownian motion. We consider lattices featuring nondiffracting speckle-like patterns, since such lattices allow flexible control of disorder characteristics. We show that the small-scale randomly located potential spots act as a random potential that produces random forces acting on soliton and thus causing its Brownian- type motion. As a result, we find that the statistically averaged squared displacement of the soliton center grows linearly with distance. Importantly, similar soliton diffusion phenomenon may occur in properly-scaled random potentials in other types of nonlinearities, including nonlocal and saturable, in which two-dimensional solitons are stable and robust.

The starting point of analysis is the nonlinear Schrödinger equation for the dimensionless amplitude of light field $q$ describing the beam propagation in a medium with saturable nonlinearity:

$$i\frac{\partial q}{\partial \xi} = -\frac{1}{2}\left(\frac{\partial^2 q}{\partial \eta^2} + \frac{\partial^2 q}{\partial \zeta^2}\right) - Eq\frac{S|q|^2 + pR}{1 + S|q|^2 + pR}. \qquad (1)$$

Thus, the lattice and soliton beams jointly create a correction to the refractive index. In Eq. (1) $\eta, \zeta$ and $\xi$ stand for the transverse and longitudinal coordinates normalized to the beam width and the diffraction length, respectively; the parameter $E$ describes the biasing static field; $S$ represents the saturation parameter; $p$ is the lattice depth; and the function $R(\eta,\zeta)$ describes the transverse lattice profile. We assume that optical lattice $R(\eta,\zeta) \sim |q_{\rm nd}|^2$ mimics the intensity distribution in the non-diffracting pattern. We set $E = 10$ and $S = 0.1$ in Eq. (1), which correspond to typical experimental conditions.

The complex amplitude of the non-diffracting field $q_{\rm nd}$ which induces the steady-state lattice can be written via the Whittaker integral [16-20] as:

$$q_{\mathrm{nd}}(\eta,\zeta,\xi) = \exp(-ib_{\mathrm{lin}}\xi)\int_{-\pi}^{\pi} A(\phi)\exp[i(2b_{\mathrm{lin}})^{1/2}(\eta\cos\phi + \zeta\sin\phi)]d\phi, \qquad (2)$$

where $b_{\mathrm{lin}}$ is the propagation constant and $A(\phi)$ is the angular spectrum. Different angular spectra generate various types of nondiffracting patterns. Thus, for $A(\phi) = \exp(im\phi)$ one gets a $m$-th order Bessel lattice [16,17] supporting different kinds of solitons [21-23]. A random $A(\phi)$ spectrum yields non-diffracting random patterns, as observed in [24]. In our case the complex random function $A(\phi)$ has a normal distribution at fixed $\phi$, with zero mean value $\langle A(\phi)\rangle = 0$ and unit variance $\langle |A(\phi)|^2\rangle = 1$ (the angular brackets stand for statistical averaging). If the correlation function of the angular spectral noise is Gaussian, namely $\langle A(\phi_1)A^*(\phi_2)\rangle = \exp[-(\phi_2-\phi_1)^2/\phi_{\mathrm{cor}}^2]$, then $A(\phi)$ may be represented by the Fourier series $A(\phi) = \sum_{m=-\infty}^{\infty} A_m \exp[-m^2/(2M^2)]\exp(im\phi)$, where $A_m$ is a sample from a complex normal distribution with zero mean value and unit variance, and the parameter $M$ characterizes the spectral width of the noise and is related to the correlation angle $\phi_{\mathrm{cor}} = \pi/M$. This representation yields a random lattice with the complex amplitude:

$$q_{\mathrm{nd}}(\rho,\phi,\xi) = C\exp(-ib_{\mathrm{lin}}\xi)\sum_{m=-\infty}^{\infty} A_m \exp[-m^2/(2M^2)]\exp(im\phi)J_m[(2b_{\mathrm{lin}})^{1/2}\rho] \qquad (3)$$

where $\rho = (\eta^2+\zeta^2)^{1/2}$ is the radius, $J_m$ is the $m$-th order Bessel function, and $C$ is the normalization constant. Such *random non-diffracting lattice* is a superposition of non-diffracting Bessel beams with random complex weight coefficients diminishing monotonically with growing $m$. Effectively, only Bessel functions of orders $m \leq M$ contribute significantly to the random lattice profile. Such lattice exhibits two key characteristic scales. The small inner scale, or the typical radius of separate bright spots (Fig. 1), may be estimated as $L_{\mathrm{inn}} \sim \rho_0/(2b_{\mathrm{lin}})^{1/2}$, where $\rho_0 \simeq 2.44$ is the first zero of zero-order Bessel function. The outer scale may be estimated as $L_{\mathrm{out}} \simeq \rho_M/(2b_{\mathrm{lin}})^{1/2}$, where $\rho_M \approx M + 1.86M^{1/3}$ is the first zero of the $M$-th order Bessel function. The normali-zation constant $C$ in Eq. (3) is selected is such way that $L^{-2}\int_{-L/2}^{L/2}\int_{-L/2}^{L/2} R(\eta,\zeta)d\eta d\zeta = 1$, where $L \gg L_{\text{out}}$ is the size of the integration domain.

Experimentally random non-diffracting patterns can be generated using computer-generated holograms and spatial light modulators [17,20]. Such patterns can also be obtained by illumination of a narrow annular slit with random angular transmission function placed in the focal plane of a lens [24]. Note that varying the angular transmission function is equivalent to modifying the angular spectrum introduced in Eq. (2). By rotating a diffuser placed after the annular slit, one may generate different realizations of the optical lattices [Figs. 1(a) and 1(b)].

When a soliton enters such random lattices with zero incidence angles, it experiences attraction toward the strongest neighboring bright lattice spots. In shallow enough lattices, the soliton, after "elastic" scattering by the neighboring bright spot, can keep moving in the transverse plane toward the next strong lattice inhomogeneity. As a result, solitons may experience *substantial* transverse displacements *far exceeding* the transverse scales of both, the solitons and the lattice spots. Hence, *solitons may diffuse* in random lattices in analogy with Brownian particles. Note that, in contrast to solitons modeled by complete integrable systems, such particle-like soliton behavior in the setting addressed here, e.g. a two-dimensional geometry, is largely nontrivial.

Typical trajectories of soliton motion for different lattice realizations are shown in Figs. 1(c) and 1(d). The profiles of the input solitons $q(\eta,\zeta,\xi) = w(\eta,\zeta)\exp(ib\xi)$ (where $w(\eta,\zeta)$ is a real function and $b$ is a propagation constant) were obtained at $p=0$ from Eq. (1) using a relaxation method. They can be characterized with the energy flow $U = \int\int_{-\infty}^{\infty} |q|^2 \, d\eta d\zeta$. Solitons may change their direction of motion in the transverse plane many times before they are asymptotically (at $\xi \to \infty$) trapped in one of the lattice spots. Deeper lattices result in larger transverse displacements in certain range of $p$. This *diffusive regime* is strongly nonlinear and is found to occur when well-localized soliton keep their identity upon motion across the lattice, in contrast to weakly localized regimes [8].

To quantity the nature of soliton diffusion, we employed a Monte-Carlo approach, studying soliton diffusion over more than 1000 realizations of $R(\eta,\zeta)$. Input solitons with fixed energy flow $U$ were launched in the center of shallow lattices that thus do not result in substantial distortions of the soliton shape. We monitored the behavior of the average squared displacement $\rho_{\text{av}}^2 = \langle \eta_{\text{c}}^2 \rangle + \langle \zeta_{\text{c}}^2 \rangle$ of soliton center with coordinates $\eta_{\text{c}} = U^{-1} \int \int_{-\infty}^{\infty} \eta |q|^2 \, d\eta d\zeta$, $\zeta_{\text{c}} = U^{-1} \int \int_{-\infty}^{\infty} \zeta |q|^2 \, d\eta d\zeta$. Figure 2(a) shows the dependence of the average squared displacement on the propagation distance. After an initial ballistic evolution where $\rho_{\text{av}}^2 \sim \xi^2$ and the soliton displacement is still smaller than the mean distance between lattice inhomogeneities, the movement of solitons in the transverse plane becomes analogous to the random walk. Thus, one of the surprising results of this work is that there exists the regime where specific nondiffracting lattice causes random soliton displacements fully analogous to that experienced by Brownian particles.

The analogy is confirmed by the results of Fig. 2(b), which shows the average squared displacement at $\xi = 20$ as a function of $p$. Increase of the lattice depth causes a linear growth of $\rho_{\text{av}}^2$ in shallow lattices, which means that the *lattice depth $p$ is analogous to the diffusion coefficient*. The analogy cannot be rigorously complete, because solitons moving across the lattice do slowly radiate and are eventually trapped by one of the lattice spots as $\xi \to \infty$. Since the soliton energy leakage and the probability of soliton trapping increase with the lattice depth $p$, the dependence $\rho_{\text{av}}^2(p)$ deviates from linear one for large enough $p$. The average squared displacement gradually saturates. Notice that excessively high lattice depths cause break-up of the input beam into several trapped fragments, thus entering a drastically different regime that the one addressed here.

The topological structure of the non-diffracting pattern strongly depends on the number of Bessel beams $M$ effectively taking part in the lattice formation [Figs. 3(a) and 3(b)]. Lattices corresponding to small and moderate $M$ values are irregular in their centers, but become almost regular at the periphery. The domain of irregularity rapidly expands with increase of $M$, so that lattices with $M \sim 20$ feature strongly irregular shapes [Figs. 1(a) and 1(b)]. When $M \to 1$, the lattice exhibits an intensity distribution similar to that of a single Bessel beam. Importantly, we found that even for moderate $M$ values, solitons *do not penetrate* into regions where lattice (completely irregular in its center) features *quasi-regular* structure. This suggests that $\rho_{\text{av}}^2$ rapidly increases with growth of $M$ or, equivalently, with increasing the width of the irregularity domain [Fig. 4(a)]. The rapid growth of $\rho_{\text{av}}^2$ is eventually replaced by saturation for $M \to \infty$. In this limit the asymptotic $\rho_{\text{av}}^2$ value is determined by the lattice depth and by the propagation distance.

We found that $\rho_{\mathrm{av}}^2$ strongly depends on the ratio between the soliton width and the inner lattice scale. A growth of $b_{\mathrm{lin}}$ causes a reduction of $L_{\mathrm{inn}} \simeq \rho_0/(2b_{\mathrm{lin}})^{1/2}$. Small-scale lattices do not lead to substantial soliton displacements because the forces acting on soliton diminish with the increasing number of inhomogeneities that it covers. Consequently, we found that $\rho_{\mathrm{av}}^2$ monotonically decreases with increase of $b_{\mathrm{lin}}$ (Fig. 4(b)). Soliton is more sensitive with respect to the refractive index fluctuations possessing the correlation length close to the soliton width. The soliton smoothes over small-scale fluctuations and it moves more slowly in the large-scale random fields because of the diminishing of the refractive index gradients. Surprisingly, we observed that the average displacement is a non-monotonic function of the energy flow of the input soliton [Fig. 4(c)]. The energy flow defines the soliton width as well as its nonlinear contribution to the refractive index. Thus the averaged squared displacement reaches its maximum when the soliton width is comparable with the correlation length, and soliton contribution to the refractive index does not exceeds remarkably typical depth of the random refractive index landscape. In the limit $U \to \infty$ soliton becomes wide (due to the saturation), and its contribution to the refractive index dominates the random one thus suppressing the Brownian-type motion. In the low-energy limit the soliton is also wide in comparison with the correlation length and Brownian-type motion is suppressed again.

When solitons are launched with a nonzero input angle $\alpha$, they experience scattering on the random inhomogeneities rather than diffusion [Fig. 3(c)]. We studied the impact of the input angle on the average squared displacement defined as $\rho_{\mathrm{av}}^2 = \langle (\eta_c - \alpha\xi)^2 \rangle + \langle \zeta_c^2 \rangle$. The average squared displacement is found to be a monotonically decreasing function of $\alpha$ [Fig. 4(d)]. In the framework of the soliton-particle analogy, increasing the input angle corresponds to growing initial kinetic energy; thus, the shallow fluctuations of the random potential become less relevant and only the major (and less-probable) ones produce significant scattering that finally results in decreasing of the averaged squared displacement. For large enough $\alpha$, $\rho_{\mathrm{av}}^2$ was found to become smaller than the displacement for solitons that are initially at rest and experience usual diffusion. Notice that soliton scattering is anisotropic. In shallow lattices the transverse scattering (along $\zeta$ axis) dominates, while in deep lattices $\eta$-displacements become more pronounced because of the growing probability of soliton trapping.

Summarizing, we have revealed a fundamental analogy between soliton diffusion in random optical lattices and Brownian motion. The analogy is substantiated by the behavior of the average squared soliton displacement, which was found to increase linearly with distance. We also predict that the average displacement grows linearly with the lattice depth in shallow lattices, a property that might be employed in practice to control the soliton diffusion in reconfigurable optically-induced lattices.

# Figure captions

Figure 1. Panels (a) and (b) show two different lattice realizations at $M = 20$ and $b_{\text{lin}} = 4$. Panels (c) and (d) show soliton diffusion in lattices with $p = 0.005$ and $p = 0.01$, respectively. Output intensity distributions at $\xi = 20$ are shown for three representative lattice realizations. White lines show trajectory of soliton center motion in the transverse plane.

Figure 2. (a) Average squared displacement of soliton center versus propagation distance for $p = 0.011$ (1) and $0.004$ (2). (b) Average squared displacement at $\xi = 20$ versus lattice depth. In all cases $U = 8.67$, $b_{\text{lin}} = 4$, and $M = 20$.

Figure 3. Representative realizations of optical lattices corresponding to $M = 4$ (a) and 8 (b). Panel (c) show diffusion of solitons with $U = 8.67$ launched at an angle $\alpha = 0.5$ in lattices with $p = 0.01$, $M = 40$. Output intensity distributions at $\xi = 20$ are shown for three representative lattice realizations. White lines show trajectory of soliton center motion in the transverse plane. In all cases $b_{\text{lin}} = 4$.

Figure 4. (a) Average squared displacement of soliton center versus parameter $M$ at $p = 0.01$, $U = 8.67$, and $b_{\text{lin}} = 4$. (b) Average squared displacement versus parameter $b_{\text{lin}}$ at $p = 0.01$, $U = 8.67$, and $M = 20$. (c) Average squared displacement versus energy flow of input soliton at $p = 0.01$, $M = 20$, $b_{\text{lin}} = 4$. (d) Average squared displacement versus input angle at $U = 8.67$, $p = 0.01$, $M = 40$, and $b_{\text{lin}} = 4$.

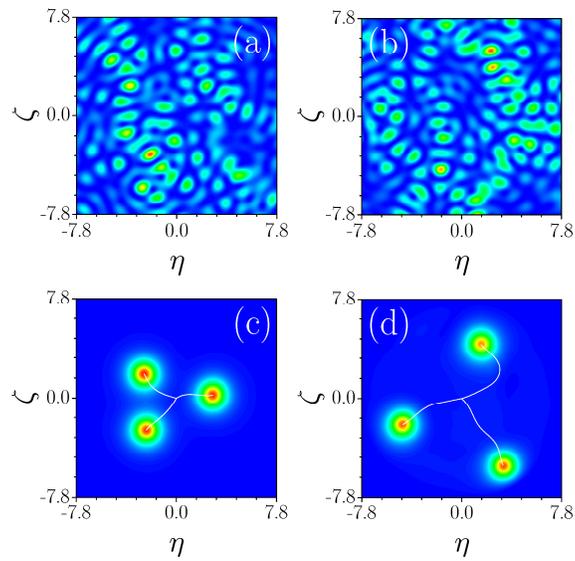

Figure 1. Panels (a) and (b) show two different lattice realizations at $M = 20$ and $b_{\text{lin}} = 4$. Panels (c) and (d) show soliton diffusion in lattices with $p = 0.005$ and $p = 0.01$, respectively. Output intensity distributions at $\xi = 20$ are shown for three representative lattice realizations. White lines show trajectory of soliton center motion in the transverse plane.

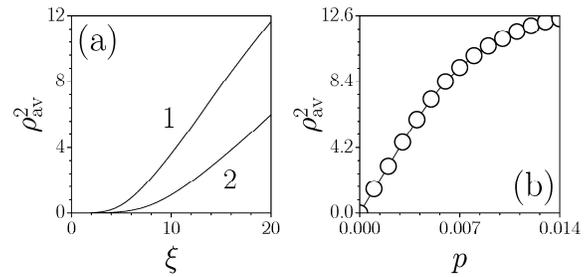

Figure 2. (a) Average squared displacement of soliton center versus propagation distance for $p = 0.011$ (1) and $0.004$ (2). (b) Average squared displacement at $\xi = 20$ versus lattice depth. In all cases $U = 8.67$, $b_{\text{lin}} = 4$, and $M = 20$.

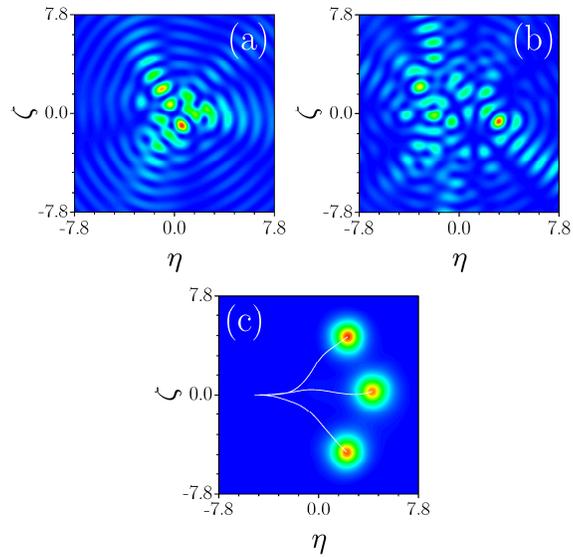

Figure 3. Representative realizations of optical lattices corresponding to $M = 4$ (a) and 8 (b). Panel (c) show diffusion of solitons with $U = 8.67$ launched at an angle $\alpha = 0.5$ in lattices with $p = 0.01$, $M = 40$. Output intensity distributions at $\xi = 20$ are shown for three representative lattice realizations. White lines show trajectory of soliton center motion in the transverse plane. In all cases $b_{\text{lin}} = 4$.

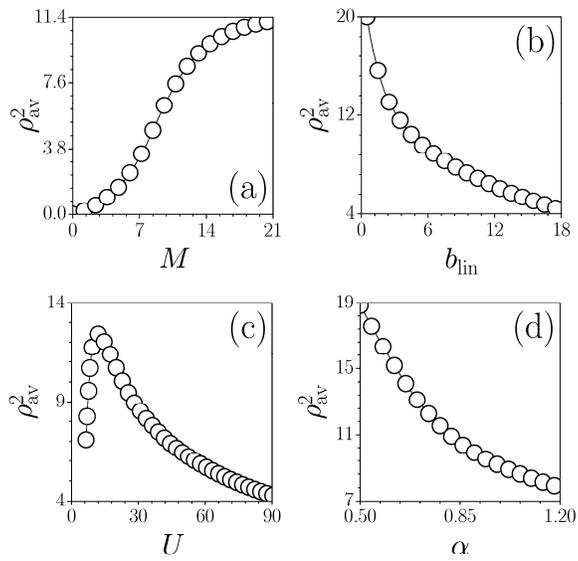

Figure 4. (a) Average squared displacement of soliton center versus parameter $M$ at $p = 0.01$, $U = 8.67$, and $b_{\text{lin}} = 4$. (b) Average squared displacement versus parameter $b_{\text{lin}}$ at $p = 0.01$, $U = 8.67$, and $M = 20$. (c) Average squared displacement versus energy flow of input soliton at $p = 0.01$, $M = 20$, $b_{\text{lin}} = 4$. (d) Average squared displacement versus input angle at $U = 8.67$, $p = 0.01$, $M = 40$, and $b_{\text{lin}} = 4$.